\documentclass[prl,twocolumn,amsmath,amssymb]{revtex4-2}
\usepackage[dvipsnames]{xcolor}
\definecolor{lgray}{gray}{0.35}
\usepackage[colorlinks=true,urlcolor=lgray,linkcolor=lgray,
citecolor=lgray,pdfpagelabels=true,hypertexnames=true,
plainpages=false,naturalnames=false]{hyperref}
\usepackage{graphicx}

\newcommand{\arXiv}[1]{\href{http://www.arXiv.org/abs/#1}{arXiv:#1}}
\renewcommand{\doi}[2]{\href{http://dx.doi.org/#2}{#1}}
\newcommand{\be}{\begin{equation}}
\newcommand{\ee}{\end{equation}}
\newcommand{\del}{\partial}
\def\nn{\nonumber}
\def\bea{\begin{eqnarray}}
\def\eea{\end{eqnarray}}
\newcommand{\dsty}{\displaystyle}
\newcommand{\de}{{\rm d}}

\begin{document}

\title{De Sitter bubbles from anti-de Sitter f{l}uctuations\vspace{-1mm}}

\author{Anxo Biasi\,}
\email{anxo.biasi@gmail.com }
\affiliation{Laboratoire de Physique de l'Ecole Normale Sup\'erieure ENS Universit\'e PSL, CNRS, Sorbonne Universit\'e, Universit\'e de Paris, F-75005 Paris, France}
\affiliation{Institute of Theoretical Physics, Jagiellonian University, Krak\'ow 30-348, Poland\vspace{-3.5mm}}
\author{Oleg Evnin\,}
\email{oleg.evnin@gmail.com}
\affiliation{Department of Physics, Faculty of Science, Chulalongkorn University, Bangkok 10330, Thailand}
\affiliation{Theoretische Natuurkunde, Vrije Universiteit Brussel and International Solvay Institutes, Brussels 1050, Belgium\vspace{-3.5mm}}
\author{Spyros Sypsas\,} 
\email{s.sypsas@gmail.com }
\affiliation{Department of Physics, Faculty of Science, Chulalongkorn University, Bangkok 10330, Thailand}
\affiliation{NARIT, Don Kaeo, Mae Rim, Chiang Mai 50180, Thailand}

\begin{abstract}
Cosmological acceleration is difficult to accommodate in theories of fundamental interactions involving supergravity and superstrings. An alternative is that the acceleration is not universal but happens in a large localized region, which is possible in theories admitting regular black holes with de Sitter-like interiors. We considerably strengthen this scenario by placing it in a global anti-de Sitter background, where the formation of `de Sitter bubbles' will be enhanced by mechanisms analogous to the Bizo\'n-Rostworowski instability in general relativity. This opens an arena for discussing the production of multiple accelerating universes from anti-de Sitter fluctuations. We demonstrate such collapse enhancement by explicit numerical work in the context of a simple two-dimensional dilaton-gravity model that mimics the spherically symmetric sector of higher-dimensional gravities.

\end{abstract}

\maketitle

Assuming that we inhabit a typical, unremarkable place in the Universe may give one a gratifying sense of fairness. Yet this fairness often incurs sharp intellectual costs.

One such example is the case of cosmic acceleration \cite{accl}. As our local cosmological neighborhood exhibits an accelerated expansion \cite{accl-obs,accl-obs-2}, extrapolating this trend to the entire universe leads to the conclusion that
we live in an approximately de Sitter (dS) spacetime. At the same time, dS solutions are difficult to accommodate in many approaches to fundamental interactions, in particular, those involving supersymmetry and superstrings \cite{dSstring1,Danielsson:2018ztv}. To quote a representative opinion \cite{dSstring2}, ``to construct a model of de Sitter space and dark energy in string theory is a great challenge.''

A possible alternative resolution of this tension is to assume that the observable universe is contained in a large but localized patch inside a much bigger universe with rather different properties. The concept of a universe-inside-a-black-hole has repeatedly resurfaced in the cosmological literature of recent decades \cite{FGG,FMM,inside1,inside2,Smolin,inside3,inside4,inside5}. Regular black holes (BHs) with dS-like interiors have been commonly considered in conjunction with gravitational singularity resolution \cite{PI,Hayward,inside3,inside4,gravastar,BHrev1,BHrev2}. A giant BH of this sort could potentially provide a dS-like environment resembling the observable universe.

Our goal in this Letter is to bring an essential new ingredient into the scenario of dynamical formation of approximately dS patches via gravitational collapse: in contrast to the past efforts framed in a Minkowski background, our ambient spacetime will be anti-de Sitter (AdS), which leads to significant new effects. Unlike dS, AdS spacetimes are straightforwardly accommodated as solutions in the currently popular theories of fundamental interactions, as evidenced, for instance, by the huge volume of research on the AdS/CFT correspondence \cite{AdSCFT}. What is even more crucial in our context is the property of AdS to enhance collapse phenomena, the Bizo\'n-Rostworowski (BR) turbulent instability \cite{BR}.

The essence of BR instability in AdS is most easily visualized by contrasting the dynamics of gravitating matter in AdS with that in Minkowski space. There are certainly collapsing matter configurations in Minkowski (as in the scalar field collapse of the classic paper \cite{Choptuik}). Yet, if one decreases the density of the initial matter configuration, there is always a threshold below which no BHs are formed and the matter gradually disperses to infinity, which underlies the stability of Minkowski space \cite{stab}. The picture of collapse phenomena in AdS is very different. Probe fields (or test particles) in Minkowski disperse out to infinity. In AdS, they are perfectly refocused by the gravitational field to the initial configuration after a certain fixed period, making the AdS space act like a perfect lens \cite{nonrel}. As a result, instead of dispersing to infinity, a matter distribution that has failed to form a BH will not expand indefinitely but reconverge and attempt making a BH again, potentially an infinite number of times. Furthermore, when dealing with gravitating fields (and not probe fields or test particles), each refocusing induced by the AdS geometry is accompanied by extra compression of the matter due to gravitational attraction, making BH formation more and more likely.

The above heuristic picture is corroborated by numerical simulations starting from \cite{BR}, where formation of BHs for Einstein's equations with a negative cosmological constant was observed for progressively smaller initial perturbations, leading to a surge of research on AdS instability (see \cite{lecture, resrev} for reviews). The position-space picture of refocusing and gradual compression, followed by collapse, is complemented by the mode-space picture of resonant turbulent transfer of energy to shorter wavelengths \cite{res1,res23,res4}. The sharp BR instability conjecture that BH formation persists in AdS down to arbitrarily small initial perturbation amplitudes remains unproved, though considerable amount of numerical evidence has accumulated. In our present context, the behavior of arbitrarily small perturbations is not so important, but the key physical effect that our subsequent considerations will build upon is that AdS enhances BH production due to its refocusing properties (in our case, regular BHs with dS-like interiors), compared to the threshold below which no BHs form in asymptotically Minkowski spaces \cite{Choptuik}.

The original considerations of \cite{BR} were set up within General Relativity (GR), where the endpoint of the BR instability is formation of AdS-Schwarzschild BHs. Our present goal is to transpose these considerations to dynamical theories where formation of regular BHs with dS-like interiors is possible, and embed them in the qualitative picture of spontaneous `dS bubble' formation from AdS fluctuations.

Because the mechanisms of the BR instability are rather robust (resting upon the refocusing properties of AdS and the attractive quality of gravitational interactions, at least at large distances), we expect that the scenario we posit here will operate in a broad class of gravity theories that have three properties:
(1) they admit vacuum AdS solutions, (2) they admit regular asymptotically AdS BHs with dS-like interiors, (3) the dynamical collapse problem is well-posed.

Static compact objects with dS-like interiors have been considered in a number of contexts.
We mention the prominent line of research involving this sort of solutions for nonlinear electrodynamics coupled to GR \cite{nlin1,nlin2,nlin3,nlin4}. A related concept is gravastars \cite{gravastar}. While it would likely not pose substantial extra difficulty to accommodate these constructions within an asymptotically AdS spacetime, they lead to complications for our pursuits since it is not straightforward to formulate a dynamical collapse problem giving birth to these objects. In nonlinear electrodynamics, constructing the electrically charged regular BHs that we need requires double-valued Lagrangians \cite{nlin1,nlin3}, which is cumbersome in a dynamical theory, while their magnetic counterparts \cite{nlin2} would require handling dynamical evolution of magnetically charged matter. It is similarly not straightforward to make gravastars emerge from a dynamical collapse process. While all of these problems are likely not insurmountable, and a variety of other constructions of regular BHs may be considered, it would be worthwhile at this point to come up with a simple, explicit setting that demonstrates that our ideas work, while keeping in mind that we expect the scenario to be robust and generic. The practical purpose of this Letter is precisely to spell out such a setting and give a practical demonstration of collapse enhancement in AdS leading to dS bubble formation.

As a test bed to validate our ideas, we choose the general two-dimensional dilaton gravity theory \cite{dilgrav1,ft-JT} of the form
\be \label{sbulk}
S_{g} = \frac12 \int \de t\,\de r \, \sqrt g 
\left( XR - U(X)(\del X)^2 -2V(X)\right),
\ee
where $X$ is the dilaton and $R$ is the 2d Ricci scalar. We couple this theory to a 2d massless scalar field $\Psi$ as~\cite{2deff1}
\be \label{S-m}
S_m = -\int \de t\,\de r \, \sqrt g \, h^2(X)
(\del \Psi)^2.
\ee
At this point, $U$, $V$ and $h$ are unspecified functions of the dilaton $X$, which we shall fix later to suit our goals~\cite{ft-U}. An important aspect for our applications of this theory is that, by tuning these functions, one may make the theory accommodate both AdS$_2$ solutions and regular BHs, while the dynamical problem for collapse of $\Psi$ is well-posed. This two-dimensional model thus provides the simplest setting to explore our ideas. What is equally important is that effective theories of the form (\ref{sbulk}-\ref{S-m}) with $r\ge 0$ emerge from the reduction of higher-dimensional gravities under the assumption of spherical symmetry
 \cite{dilgrav1,2deff1}, leading to the name `spherically reduced gravity.'
For the reduction from 4d, one would use the spherically symmetric ansatz
\bea\label{metric4d}
\de s^2= g_{tt}(t,r)\,\de t^2&+2 g_{tr}(t,r)\, \de t \,\de r +g_{rr}(t,r)\,\de r^2 \nn\\
&+h^2(X(t,r)) \,\de\Omega^2,
\eea
where $\de\Omega^2$ is the line element of a 2-sphere. Analogous reduction is, of course, possible starting from higher-dimensional theories. Table~1 of \cite{dilgrav2} gives a long list of assignments of $U$ and $V$ resulting from such spherical reductions of various gravitational theories. Note in particular that AdS$_4$ fits into (\ref{metric4d}) when represented as
\be
\de s^2= -\left(1+k^2r^2\right) \de t^2 + \frac{1}{1+k^2r^2} \de r^2+r^2 \de \Omega^2,
\ee
and so do its spherically symmetric perturbations. The 2d metric corresponding to AdS$_4$ is precisely AdS$_2$. Thus, even though we talk about collapse in AdS$_2$ in the theory (\ref{sbulk}-\ref{S-m}), there is a contact with the spherically symmetric collapse problem in higher-dimensional AdS.

The equations of motion of the action $S_g+S_m$ are
\bea
&\nabla_\mu \partial_\nu X - g_{\mu\nu} \nabla^2 X - g_{\mu\nu} V + U \partial_\mu X \partial_\nu X -\frac12 g_{\mu\nu} U(\partial X)^2\nn \\&\hspace{25mm}= -2 h^2 \partial_\mu \Psi \partial_\nu \Psi +g_{\mu\nu} h^2 (\partial \Psi)^2,\nn 
\\
&R+ \partial_X U (\partial X)^2 + 2U \nabla^2 X -2\partial_X V = 4 h \partial_X h (\partial \Psi)^2\!\!,\nn
\\
&\nabla_\mu \left( h^2(X) \partial^\mu \Psi \right) =0.\nn
\eea
These simplify if one chooses to define the $r$-coordinate so that $X(t,r)$ is a prescribed function of $r$,
\be\label{dotX0}
\dot X=0.
\ee
Thereafter, we can still use redefinitions of $t$ to set the $g_{tr}$ component of the metric to 0, and parametrize the metric as \cite{2deff2}
\be \label{da-metric-Kunst}
\de s^2 = e^{2\rho(t,r)} \left(-\sigma^2(t,r) \,\de t^2 + \de r^2 \right).
\ee
The independent equations of motion can be written as
\bea\label{rhoeq}
\begin{split}
&\displaystyle X'' - \rho' X' + e^{2\rho}V +\frac12 U {X'}^2 =  -h^2 \left( \frac{\dot\Psi^2}{\sigma^2} + {\Psi'}^2 \right)\!,\hspace{9mm}
\\
&\displaystyle X''+  \frac{\sigma'}{\sigma}X'  = -2 e^{2\rho} V(X),\quad
 \left({h^2\dot\Psi}/{\sigma}\right)^{\hspace{-.5mm}\displaystyle\mathbf{.}}=\left(h^2\sigma \Psi'\right)'.
\end{split}
\eea
One has additionally an equation of the form
$ \dot\rho X' = 2 h^2 \dot\Psi \Psi'$,
which only needs to be enforced at one spatial point, since its $r$-derivative is a consequence of (\ref{rhoeq}).
The collapse problem is well-posed in the context of (\ref{rhoeq}), but before we proceed with solving it, we must decide on the choice of functions $U$, $V$ and $h$, guided by the properties of vacuum BH solutions in this theory.

In vacuo, $\Psi=0$ and (\ref{rhoeq}) is simplified to
$$
X'' - \rho' X'  +\frac12 U {X'}^2 =  - e^{2\rho}V,\quad
X''+  \frac{\sigma'}{\sigma}X'  = -2 e^{2\rho} V.
$$
We make a concrete choice for the definition of $r$, setting $X$ to be a time-independent function satisfying 
$X''=-UX'^2$,
which can be integrated to
\be
X'=e^{-Q(X)},\qquad Q(X) \equiv Q_0 + \int^X \de y \; U(y).\label{Xprimegauge}
\ee
In this gauge, the equations of motion are solved by
\be
\sigma e^{2\rho}=1,\qquad \sigma e^{-Q}= -2M+w,
\ee
where $M$ is an integration constant playing the role of the BH mass, with
\be
w \equiv -2 \int \de r\,V(X(r)) =- 2\int^X \de y \;e^{Q(y)} V(y) . \label{wdef}
\ee
The metric is then of the form
\be \label{g}
\de s^2 = -\xi(r) \de t^2 + \frac{1}{\xi(r)} \de r^2,
\ee
where $\xi$ as a function of the dilaton is given by
\be \label{xiX}
\xi(X) = e^{Q(X)} \left( w(X)-2M \right).
\ee
The dilaton is recovered from
(\ref{Xprimegauge}).
These vacuum solutions agree with \cite{dilgrav1,dilgrav3}.

We want to have regular asymptotically AdS BHs with approximately dS interiors as vacuum solutions of our theory. The vacuum solutions are described by (\ref{g}-\ref{xiX}), with $M$ controlling the BH mass. The $M=0$ solution must thus agree with AdS$_2$, 
\be \label{AdS2}
\de s^2 = -\left(1+k^2r^2\right) \de t^2 + \frac{1}{1+k^2r^2} \de r^2,
\ee
hence
\be \label{ads-cond2}
 e^{Q(X(r))} w(X(r)) = 1+k^2r^2.
\ee
Thereafter, the general vacuum solution (\ref{g}) comes with
\be\label{xigen}
\xi(r)=1+k^2r^2-2Me^Q,
\ee
which indeed looks like an asymptotically AdS BH metric, with the precise shape of the BH controlled by the function $e^Q$. 
Following \cite{2deff1}, we focus \cite{Bardeen} on Poisson-Israel (PI) regular BHs \cite{PI,Hayward},
setting
\be
e^{Q(r)} \equiv \frac{1}{X'}= \frac{r^2}{r^3+\mu^3},\label{eQ_IP}
\ee
with $\mu$ some dimensionful parameter.
From (\ref{Xprimegauge}), (\ref{wdef}), (\ref{ads-cond2}) and (\ref{eQ_IP}),
\bea \label{V-IP}
V(r)&=&- \frac12 \left(1 + 3 k^2 r^2 - \frac{2\mu^3}{ r^3}\right) , 
\\ \label{U-IP}
U(r) &=& 
-r \frac{r^3 - 2 \mu^3 }{(r^3 + \mu^3)^2},\quad
X(r) =  \frac{r^2}2 - \frac{\mu^3}r .
\eea
Then, (\ref{V-IP}-\ref{U-IP}) implicitly define the functions $U(X)$ and $V(X)$ that yield AdS-PI regular BHs. In practice, we will not need anything other than the explicit expressions (\ref{eQ_IP}) and (\ref{V-IP}) to write down the equations of motion.

Note that the expansion of $\xi$ in (\ref{g}) near the origin for AdS-PI BHs is of the form
\be\label{effLambda}
\xi(r)=1+(k^2-2M/\mu^3)r^2+\ldots \;.
\ee
The value of $k^2$, directly related to the cosmological constant, thus gets effectively adjusted. If $M$ is sufficiently large, its sign is flipped, leading to a region of approximately dS spacetime. Note that the dS metric is of the form (\ref{AdS2}), except that one must replace $k^2$ by $-k^2$. This observation underlies the claim that PI BHs possess dS interiors. A more detailed discussion of the inner structure of PI BHs may be found in \cite{2deff1}.

Regarding the choice of $h^2(X)$, it is natural to fix it in such a way that, with $X(r)$ corresponding to static solutions substituted, it becomes simply $r^2$, so that the 4d metric (\ref{metric4d}) becomes a conventional regular BH metric. Thus, we define it by $h^2(X)=r^2(X)$, where $r(X)$ is the inverse of the dilaton configuration (\ref{U-IP}).

With $U$, $V$ and $h$ fixed as above, the equations of motion for the gravitational field become
\be\label{primeeq2}
\begin{split}
(e^{2\rho}\sigma)'X'&=2e^{2\rho}\sigma r^2 \left( \Pi^2 + {\Psi'}^2 \right),\\
(\sigma X')'&=-2(e^{2\rho}\sigma) V(r),
\end{split}
\ee
where we have introduced the momentum $\Pi\equiv {\dot\Psi}/{\sigma}$.
One can rewrite these in a slightly more convenient form introducing $e^{2\omega}\equiv e^{2\rho}\sigma$.
Then, using the explicit expressions (\ref{eQ_IP}) and (\ref{V-IP}) for the AdS-PI BHs, we get
\be\label{primeeqPI}
\begin{split}
&\hspace{5mm}\omega'=\frac{r^4}{r^3+\mu^3} \left( \Pi^2 + {\Psi'}^2 \right),\\
&\left(\frac{r^3+\mu^3}{r^2}\sigma\right)'= \left(1 + 3 k^2 r^2 - \frac{2\mu^3}{ r^3}\right)e^{2\omega},
\end{split}
\ee
while the equations for $\Psi$ originating from (\ref{rhoeq}) are
\be \label{eq:Pidot_Psidot}
\dot\Pi=r^{-2}\left(r^2\sigma \Psi'\right)',\qquad \dot\Psi=\sigma\Pi.
\ee
As a cross-check, under the identification $k=\mu=0$, $e^{2\omega}\sigma\equiv\alpha^2$, $e^{2\omega}/\sigma\equiv a^2$, these equations reduce to the equations of motion of \cite{Choptuik}, up to the irrelevant overall normalization of $\Psi$. Similarly, under the identification $\mu=0$, $k=1$, $r=\tan x$, $\delta=-2\omega$, $A=\sigma e^{-2\omega} \cos^2x$, we are in agreement with the equations of motion of \cite{BR}.

For the integration of (\ref{primeeqPI}-\ref{eq:Pidot_Psidot}), one first specifies the initial values of $\Pi$ and $\Psi$, then the metric functions $\omega$ and $\sigma$ are obtained by the integration of (\ref{primeeqPI}). The values of $\Pi$ and $\Psi$ can then be updated by (\ref{eq:Pidot_Psidot}) to the next time slice. For numerics, it is convenient to use a compact spatial coordinate $x\in[0,\pi/2)$, so that $r=\tan{x}$. The formation of the apparent horizon of the regular BH is signaled by the drop of $\sigma$ to zero at some point, denoted by $x_H$. The numerical integration scheme, which generally follows the guidelines of \cite{lecture}, is described in the Appendix.

\begin{figure}[t!]
	\centering	
	\includegraphics[width=\columnwidth]{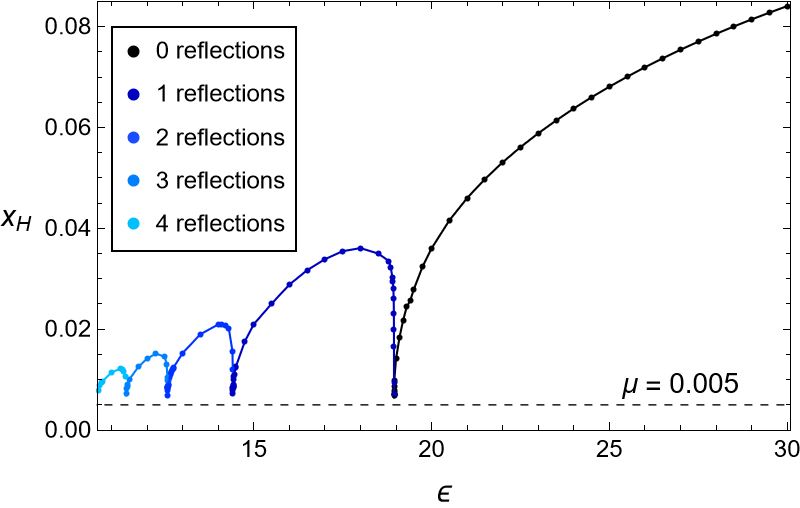}
	\vspace{-0.8cm}
	\caption{Apparent horizon radius vs. the amplitude of the initial data (\ref{eq:initial_data}). Each `hump' is associated with a given number of reflections from the AdS boundary, 0, 1, 2, 3 from right to left. Collapse in Minkowski would not have happened below the edge of the rightmost hump, but it does happen in AdS. (In the Appendix, we give for comparison the corresponding plot for pure GR in asymptotically AdS and Minkowski situations.)}
	\label{fig:Xh_vs_epsilon}
\end{figure}

Simulations of smooth initial data evolution subject to the Dirichlet boundary condition $\Psi(t,\pi/2) = 0$ demonstrate the enhancement of BH formation by the action of AdS refocusing. We have fixed $k=1$, $\mu=0.005$, and simulated a family of Gaussians
\be \label{eq:initial_data}
\hspace{-1.8mm}\Psi(0,x) = 0, \,\,\,\,\,\Pi(0,x) = \epsilon \cos^2 x \exp\left[- (10\tan x)^2\right].
\ee
Note that the choice $k=1$ simply corresponds to measuring lengths in units of the AdS radius. Thereafter, $\mu$ specifies the fraction of the AdS radius at which deviations of the AdS-PI BH metric from Schwarzschild become significant.
For $\epsilon$ large enough, there is an immediate collapse into a regular BH. For smaller $\epsilon$, no BH is formed during the first wavefront convergence, and the scalar waves expand out to the AdS boundary. They are however reflected and reconverge to the origin to form a BH on the second attempt. With $\epsilon$ decreasing further, the collapse is delayed until after two, three, four, etc reflections. A BH is thus formed for values of $\epsilon$ that would be too small in Minkowski space, by virtue of refocusing properties of AdS, in direct parallel to the original BR instability \cite{BR}. (In a context where BHs form from statistical fluctuations, such lowered threshold may dramatically increase the production rate.) Fig.~\ref{fig:Xh_vs_epsilon} reports the dependence of the apparent horizon radius $x_H$ at the moment of BH formation on $\epsilon$. It clearly shows that, in AdS, BH formation occurs even in those cases where the initial data amplitude is too small to form a BH at the first attempt.
\\
\indent Since AdS acts like a cavity, if a horizon has formed, eventually, all the available matter will end up inside the emerging BH.
For static AdS-PI BHs corresponding to the total mass of the initial data in the simulations of Fig.~\ref{fig:Xh_vs_epsilon}, the sign of the effective cosmological constant at the center of the AdS-PI BH given by (\ref{effLambda}) indeed gets flipped.

We have thus demonstrated that collapse enhancement in AdS via mechanisms related to the BR instability occurs in theories admitting regular BHs with dS-like interiors.
The coordinates used for our simulations do not allow for probing the evolution after the horizon formation,
which would require using instead adaptive coordinates that coincide with ours while the wavefronts oscillate in AdS, but become infalling once the BH has started to form. This remains an interesting outstanding problem, even in pure GR. Simpler analogous simulations in asymptotically Minkowski spacetimes have been performed in \cite{2deff1} and demonstrated that the collapsing matter piles up at the inner horizon of the regular BH it has produced, as a manifestation of the `mass inflation' effect \cite{massinf1,massinf2}. One may expect that, after the horizon has formed, the remnant will settle to a BH configuration of the theory given by (\ref{g}) and (\ref{xigen}-\ref{eQ_IP}), possibly in a variation of the present dynamical setup. A picture that we find particularly attractive has been developed in \cite{massinf1}: as the Hawking radiation carries away the residual matter, the collapse remnant settles to a vacuum extremal PI BH with a dS core.

Our considerations within the effective collapse theory (\ref{sbulk}-\ref{S-m}) provide a blueprint for what one should expect for the spherically symmetric sector of higher-dimensional gravitational theories that support regular BHs with approximately dS interiors. One should look for such theories admitting a dynamical collapse formulation, either by refining the setup of \cite{nlin1,nlin2,nlin3,nlin4} or by constructing other realizations of static BHs with dS cores.

Evidently, if one is aiming at a genuine cosmological setting, the spherical symmetry assumption should be relaxed. It is generally believed that the BR instability in GR is not restricted to the spherically symmetric sector (see \cite{nonspher1} for an early discussion, and also \cite{squash1,squash2}). Part of the reason is that the `perfect lens' properties of AdS by no means depend on spherical symmetry. A similar belief is then natural for AdS collapse enhancement in gravitational theories with regular BHs. State-of-the-art numerical simulations in GR have reached the capacity in recent years to track the evolution of non-spherically-symmetric perturbations in AdS, observing an even stronger collapse enhancement due to AdS effects \cite{nonspher2} than what had been seen in spherically symmetric simulations.

Going a step further, one may envisage spontaneous creation of multiple dS bubbles from AdS fluctuations. While dynamically this setup is much more complicated than the spherically symmetric collapse with prescribed initial data that we have treated here at the technical level, it is natural to expect that, as long as the AdS fluctuations remain small almost everywhere, they will not substantially upset the `perfect lens' properties of AdS that underlie the collapse enhancement. If successful, such a scenario will result in multiple dS bubble formation, with different effective values of the cosmological constant, cf. (\ref{effLambda}), providing a concrete approach to the multi-region physical realization \cite{Smolin} of anthropic arguments. To sum up,
weaving together the prominent but hitherto disjoint research trends of regular black holes and turbulent AdS instability unlocks a novel approach to accommodating
localized patches of accelerated cosmological expansion in theories without global dS solutions.

\begin{acknowledgments}
\noindent {\bf Acknowledgments:} AB has been supported by the Polish National Science Centre grant No. 2017/26/A/ST2/00530 and by the LabEx ENS-ICFP: ANR-10-LABX-0010/ANR-10-IDEX-0001-02 PSL*. OE and SS have been supported by Thailand NSRF via PMU-B (grant numbers B01F650006 and B05F650021). OE has been additionally supported by FWO-Vlaanderen through project G006918N.
\end{acknowledgments}

\appendix

\section{Appendix: Numerical integration specifics}

For the purpose of numerical simulations, it is convenient to further recast the equations of motion (\ref{primeeqPI}-\ref{eq:Pidot_Psidot}) in a slightly different form that takes them closer to the setup of \cite{BR}. First, we switch to a compact spatial coordinate $x\in[0,\pi/2)$, so that 
\be
r = \tan{x}.
\ee 
Then, after the redefinition $P(t,x) = \Pi(t,x) / \cos^2x$, $\Phi(t,x) = \partial_x\Psi(t,x)$, and $S(t,x) = \sigma(t,x)\cos^2x$, equations (\ref{primeeqPI}-\ref{eq:Pidot_Psidot}) take the form (primes will denote differentiation with respect to $x$ from now on)
\bea \label{eq:Equation_w}
\omega' &=& \frac{\cos{x} \sin^4 x}{\Delta} \left(P^2 + \Phi^2\right),\\
\label{eq:Equation_S}
S' &=& q e^{2\omega} + \frac{\nu'}{\nu}S,\\
\label{eq:Equation_P_Phi}
\dot{\Phi} &=& \left(S P\right)', \qquad \dot{P} = \frac{1}{\tan^2x}\partial_x\left(\tan^2x S \Phi\right),
\eea
with functions
\be\nn
\begin{split}
	&\Delta(x) \equiv\sin^3x + \mu^3 \cos^3x, \quad \nu(x) \equiv \frac{\cos^3x \sin^2x}{\Delta},\\
	&q(x) \equiv\frac{1}{\Delta} \left(\cos{x} \sin^2x - 2\mu^3 \frac{\cos^4x}{\sin{x}} + 3k^2 \frac{\sin^4x}{\cos{x}}\right).
\end{split}
\ee
Once $P$ and $\Phi$ are specified, the metric function $\omega$ comes from integrating (\ref{eq:Equation_w}) directly, with 
\be
\omega(t,0)=0,
\label{eq:W=0_Interior_Time_Gauge}
\ee 
to fix the time-reparametrization freedom. One can integrate (\ref{eq:Equation_S}) to obtain
\bea \label{eq:Equation_S_integral_form}
&\hspace{-2cm}\dsty S(t,x) = e^{2\omega(t,x)}\left(1+(k^2-1)\frac{\sin^5x}{\Delta}\right) \\
& \hspace{1cm}\dsty- \nu(x)\int_0^{x}e^{2\omega(t,y)} F(y) \left(P^2(t,y) + \Phi^2(t,y)\right)\de y, \nn
\eea
with 
$$
F(y)\equiv 2\frac{\sin^2{y}}{\Delta(y)} \left(\mu^3 \cos{y}+ \sin^3{y} \left(1+k^2 \tan^2y\right)\right).
$$
Note that, by itself, (\ref{eq:Equation_S}) only fixes (\ref{eq:Equation_S_integral_form}) up to a term of the form $\nu(x) c(t)$ with an arbitrary $c(t)$. The momentum constraint, which is the equation mentioned immediately under (\ref{rhoeq}), fixes $c(t)$ to be constant, and it is then set to zero by our choice to work
with initial conditions that do not already contain a PI BH.
To sum up, we numerically solve the equations (\ref{eq:Equation_w}), (\ref{eq:Equation_P_Phi}), and (\ref{eq:Equation_S_integral_form}).

For $\mu\neq0$, solutions to (\ref{eq:Equation_w}-\ref{eq:Equation_P_Phi}) near the origin $x=0$ have the structure 
\bea
&\dsty \psi(t,x) \sim \psi_0(t) + \mathcal{O}\left(x^2\right), \quad \omega(t,x) \sim \frac{\dot{\psi_0}^2}{5\mu^3} x^5 + \mathcal{O}\left(x^7\right),\nn\\
& S(t,x) \sim 1 + \mathcal{O}\left(x^2\right).
\eea
Near the boundary $x=\pi/2$, subject to reflective boundary conditions
\be
\Phi(t,\pi/2)=P(t,\pi/2)=0,
\ee
solutions have the structure ($\rho \equiv x - \pi/2$)
\bea
&\dsty\psi(t,x) \sim \psi_\infty(t) \rho^3 + \mathcal{O}\left(\rho^5\right), \quad \omega(t,x) \sim \omega_\infty(t) + \mathcal{O}\left(\rho^6\right), \vspace{1mm}\nn\\
&\dsty S(t,x) \sim k^2 e^{2\omega_\infty(t)} + \left(1-k^2\right) e^{2 \omega_\infty(t)}\rho^2 \nn\\
&\dsty + \frac{M}{6} e^{2\omega_\infty(t)} \rho^3 + \mathcal{O}\left(\rho^4\right),\nn
\eea
where $M$ is the conserved total energy inside the space, which has the form
\be \label{eq:Total_Energy}
M = e^{-2\omega(t,{\pi}/{2})}\int_0^{\frac{\pi}{2}}e^{2\omega} F(y) \left(P^2 + \Phi^2\right)\de y.
\ee

The numerical integration scheme for the equations of motion closely parallels the one described in detail in \cite{lecture} for the GR case ($\mu=0$). Specifically, a 4th-order Runge-Kutta method with variable time-step ($\Delta t = \frac{1}{3}e^{-2\omega(t,\pi/2)}\Delta x$) is used to advance in time. Spatial differentiation is performed by a 4th-order finite difference scheme on a grid of $2^{17}+1$ points. To check the validity of our simulations, we monitor the conservation of the total energy (\ref{eq:Total_Energy}) and the momentum-constraint $\dot{A} = - 4 \sin{x} e^{2\omega} A^2 P \Phi$ where $A\equiv e^{2\omega} S$. With our scheme, the total energy (\ref{eq:Total_Energy}) is conserved in the most demanding moments to an accuracy better than $10^{-8}$, and the momentum constraint is satisfied to an accuracy better than $10^{-6}$. The formation of the apparent horizon of regular BHs is signaled by a drop of $A(t,x)$ to zero at some point, denoted by $x_H$. The numerical criterion to decide whether the BH is formed in a simulation of $2^n+1$ grid-points is when the minimum of $A(t,x)$ becomes smaller than $2^{7-n}$ \cite{lecture}.

\begin{figure}[t!]
	\centering	
	\includegraphics[width=\columnwidth]{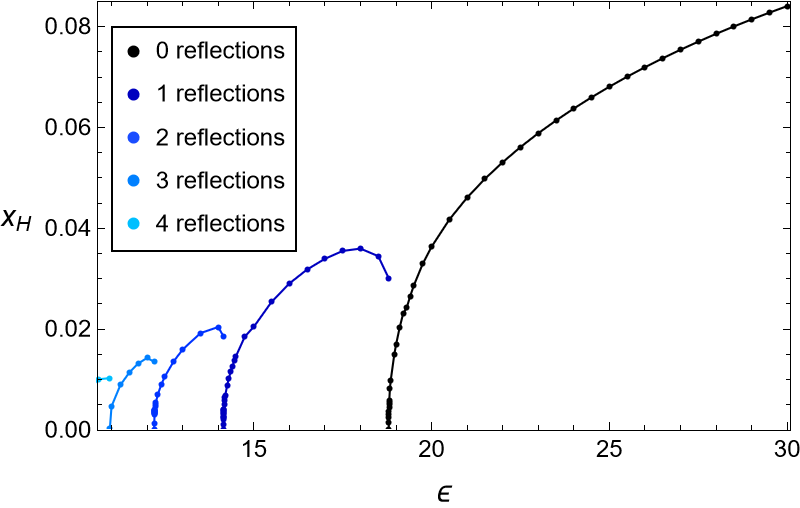}
	
	\vspace{0.1cm}
	\includegraphics[width=\columnwidth]{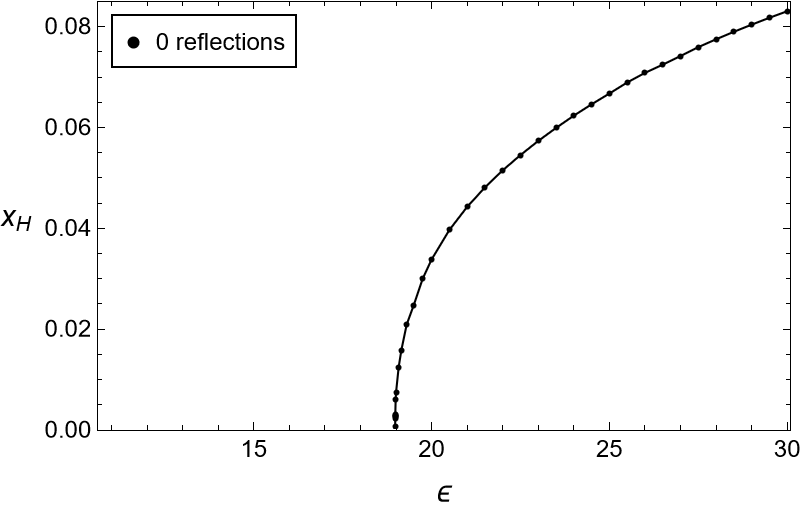}
	
	\vspace{-0.cm}
	\caption{Apparent horizon radius vs. the amplitude of the initial data given by (\ref{eq:initial_data}) for asymptotically AdS (top) and Minkowski (bottom) cases in GR, to be compared with Fig.~\ref{fig:Xh_vs_epsilon} for dynamical collapse in a theory admitting AdS-PI regular BHs. In Minkowski, BHs do not form if direct collapse fails, in contrast to the main story of AdS collapse enhancement discussed in this paper.}
	\label{fig:Xh_vs_epsilon_AdS_MK_MU}
\end{figure}

We conclude by reproducing the standard results in GR literature for scalar field collapse in asymptotically AdS spacetimes originally treated in \cite{BR} and corresponding to $k=1$, $\mu=0$ in our language,
and in asymptotically Minkowski spacetimes, treated in \cite{Choptuik} and corresponding to $k=\mu=0$ in our language. These results are presented in Fig.~\ref{fig:Xh_vs_epsilon_AdS_MK_MU} and can be contrasted with our novel numerical simulations for theories with regular BHs displayed in Fig.~\ref{fig:Xh_vs_epsilon}. The AdS-GR case is similar to the corresponding plot in Fig.~\ref{fig:Xh_vs_epsilon} (the difference is that $x_H$ may drop to zero here, while for PI BHs there is a minimal BH size below which it loses its horizons).
In Minkowski, we see that black hole formation ceases at a finite initial perturbation amplitude as a reflection of Minkowski stability \cite{stab}.
\\
\indent A curious feature of the AdS-GR case is that the horizon size $x_H$ appears to change discontinuously when the number of reflections from the AdS boundary changes, while in the corresponding plot in Fig.~\ref{fig:Xh_vs_epsilon} involving AdS-PI BHs, the curve is continuous.
A heuristic explanation can be given as follows. In a regime where collapse occurs after $n$ reflections from the AdS boundary, as we decrease the initial data amplitude, the horizon radius at the moment of BH formation will decrease until it reaches the finite value where AdS-PI BHs lose their horizons. When this happens, collapse after $n$ reflections will no longer be successful and the wavefront will expand again and then reconverge to the center. Due to the perfect lens property of AdS, the reconverging wavefront will look almost identical to the $n$th collapse attempt, except that it will be slightly more compressed due to the continued action of gravitational attraction and will actually form an AdS-PI BH on the next attempt with the horizon radius just slightly above the AdS-PI BH threshold. It is the perfect lens property of AdS that underlies the continuity.
By contrast, in standard GR, as we decrease the initial data amplitude, collapse after $n$ reflections will continue until $x_H=0$. At that point, an infinitesimal fraction of the wavepacket participates in the collapse process, so after the amplitude has been decreased further and the collapse fails, the collapse after $n+1$ reflections will have to be based on the extended part of the wavepacket that did not participate in the formation of the tiny horizon in collapse after $n$ reflections. This presumably undermines the continuity of the collapse process in GR depending on the number of reflections.


\begin{thebibliography}{99}
\bibitem{ft-U}
The kinetic potential $U$ can be shifted to zero by a Weyl rescaling. We choose to keep it in order to make the connection with higher dimensional theories more straightforward. See~\cite{dilgrav1} for a discussion on this point.
\end{thebibliography}
\end{document}